\RequirePackage{fix-cm}
\documentclass[smallextended]{svjour3}
\smartqed  
\usepackage{appendix}
\usepackage{amsmath}
\usepackage{graphicx}
\usepackage{lineno}
\usepackage{array}
\usepackage{longtable}
\usepackage{natbib}
\usepackage{subcaption}
\usepackage{soul}
\usepackage{hyperref}
\usepackage[printwatermark]{xwatermark}
\usepackage{xcolor}
\newwatermark[allpages,color=gray!25,angle=45,scale=3,xpos=-30,ypos=25]{DRAFT}

\newcommand*\patchAmsMathEnvironmentForLineno[1]{%
\expandafter\let\csname old#1\expandafter\endcsname\csname #1\endcsname
\expandafter\let\csname oldend#1\expandafter\endcsname\csname end#1\endcsname
\renewenvironment{#1}%
{\linenomath\csname old#1\endcsname}%
{\csname oldend#1\endcsname\endlinenomath}}%
\newcommand*\patchBothAmsMathEnvironmentsForLineno[1]{%
\patchAmsMathEnvironmentForLineno{#1}%
\patchAmsMathEnvironmentForLineno{#1*}}%
\AtBeginDocument{%
\patchBothAmsMathEnvironmentsForLineno{equation}%
\patchBothAmsMathEnvironmentsForLineno{align}%
\patchBothAmsMathEnvironmentsForLineno{flalign}%
\patchBothAmsMathEnvironmentsForLineno{alignat}%
\patchBothAmsMathEnvironmentsForLineno{gather}%
\patchBothAmsMathEnvironmentsForLineno{multline}%
}
\usepackage{xcolor}

\begin{document}


\title{Does the Venue of Scientific Conferences Leverage their Impact? A Large Scale study on Computer Science Conferences}

\titlerunning{Does the Venue of Scientific Conferences Leverage their Impact?}        

\author{Luca Bedogni \and Giacomo Cabri \and Riccardo Martoglia \and Francesco Poggi}
\authorrunning{Bedogni, Cabri, Martoglia, Poggi}   
%
%
\institute{
L. Bedogni, G. Cabri and R. Martoglia \at 
Dipartimento di Scienze Fisiche, Informatiche e Matematiche\\Universt\`a di Modena e Reggio Emilia, Via Campi 213/a, Modena, 41125, Italy,\\
\email{\{luca.bedogni, giacomo.cabri, riccardo.martoglia\}@unimore.it}
\and
F. Poggi \at
Dipartimento di Comunicazione ed Economia\\Universt\`a di Modena e Reggio Emilia, Viale Allegri 9, Reggio Emilia, 42121, Italy,\\
\email{francesco.poggi@unimore.it}
}

\date{Received: date / Accepted: date}

\maketitle

\begin{abstract}

\textit{Background:} Conferences bring scientists together and provide one of the most timely means for disseminating new ideas and cutting-edge works. The importance of conferences in scientific areas is testified by quantitative indicators. In Computer Science, for instance, almost two out of three papers published on Scopus are conference papers.

\textit{Objective/Purpose:} The main goal of this paper is to investigate a novel research question: is there any correlation between the impact of a scientific conference and the venue where it took place?

\textit{Approach:} In order to measure the impact of conferences we conducted a large scale analysis on the bibliographic data extracted from 3,838 Computer Science conference series and over 2.5 million papers spanning more than 30 years of  research. To quantify the "touristicity" of a venue we exploited some indicators such as the size of the Wikipedia page for the city hosting the venue and other indexes from reports of the World Economic Forum.

\textit{Results/Findings:} We  found  out  that  the  two  aspects  are  related, and the correlation with conference impact is stronger when considering country-wide touristic indicators, such as the Travel\&Tourism Competitiveness Index. Moreover the almost linear correlation with the Tourist Service Infrastructure index attests the specific importance of tourist/accommodation facilities in a given country.

\textit{Conclusions:} This is the first attempt to focus on the relationship of venue characteristics to conference papers.
The results open up new possibilities, such as allowing conference organizers and authors to estimate in advance the impact of conferences, thus supporting them in their decisions.


\end{abstract}

\vspace{8pt}
\noindent \textbf{Acknowledgements} The authors would like to thank Nicolay Osalchuk, who performed a preliminary analysis of the data during his bachelor internship.
\\
This research has been supported by the University fund for Research (FAR 2020) of the University of Modena and Reggio Emilia.

\section{Introduction}
\label{intro}


Conferences play an important role in many disciplines, in particular in scientific areas. 
Given  the  evolution  speed  of  these  fields,  conference papers are perceived as one of the most timely means for disseminating new ideas since the cycle of publication is much shorter than journal ones. Moreover, conferences open up opportunities to bring scientists together and introduce cutting-edge works, even if less mature and sophisticated. 
As a consequence in specific disciplines like Computing and Information Science, for instance, 62.3\% of the articles published on Scopus are conference papers, while only about half (i.e. 32.8\%) were journal papers, as reported in \cite{scopus2020}.

As researchers, our main goal is to provide our scientific communities with novel contributions that increase our body of knowledge or enable us to see the world under a different point of view. However our contributions must be validated and spread by appropriate means such as international conferences. 
Of course, researchers choose the most suitable venue depending on the specific contribution, but we wonder whether there are other factors that can influence the choice. There are studies that have shown that such factors exist for conferences (see Section~\ref{RelatedWork}) but no work has focused on the \emph{touristic appeal} of the conference venue. 

Our initial idea was that a conference venue able to attract tourists is also attractive for researchers to submit their paper.
Of course our claim is not that the venue is the ``only'' or ``most important'' factor, but we claim that it is a factor considered by researchers, as hosting a conference in a more touristic place also offers better experiences for  participants outside of the conference sessions.
However, as scientists, we are aware that a feeling is not enough and an idea must be supported by data and their analysis.
So, we decided to explore the correlation between the attractiveness of a conference and the ``touristicity'' of its venue from a data-driven point of view.

First of all, we had to better define the two concepts to be correlated.

Intuitively, the attractiveness of a conference is defined by the number of submissions it receives; unfortunately, this number is not so easy to retrieve: while the number of the accepted papers is public, the number of the submitted papers is not and only some conferences publish it.
So we decided to rather focus on the conference \textit{impact}, and we did it by exploiting another indicator, the number of the citations of the papers published in the conference proceedings: the two aspects go indeed hand in hand, as an attractive conference with a lot of submissions can select better papers that will have eventually more citations, therefore having a higher impact on the research community.

In the same way, the ``\emph{touristicity}'' of a venue can be very intuitive but is not straightforward to be formally defined.
To this purpose, we exploited some indicators such as the size of the Wikipedia page for the city hosting the venue, the number of annual tourist arrivals and other officially defined indexes from the Travel and Tourism competitiveness report of the World Economic Forum\footnote{https://www.weforum.org/}.

Our main research questions are therefore the following:
\begin{itemize}
    \item \textbf{RQ1}: is there any correlation between the impact of a scientific conference and the venue where it takes place?
    \item \textbf{RQ2}: if yes, which are the venue factors that are more involved in the correlation?
\end{itemize}

To answer these questions, we focused on the conferences in the field of Computer Science and Information Technology and we built a large dataset comprising nearly 4000 conference series and over 2.5 million papers spanning more than 30 years.




We grouped and aggregated the data following different criteria and producing data visualizations that enabled us to analyze the context from different points of views. We computed the required indicators from such views and, more specifically, we computed the correlation between the impact indicators and the touristicity indicators using three different correlation measures. We present the obtained results including the discussion of some representative examples, and we show that a correlation exists between the number of citations and the indexes from the Travel and Tourism competitiveness report; in particular, there are two indexes (TTCI -- Travel \& Tourism Competitiveness Index --  and TSI -- Tourist service infrastructure) which show almost a linear correlation.

This paper exploits our past expertise in bibliometric analyses (\cite{Poggi2019}, \cite{Nuzzolese2019539}, \cite{Peroni2020253}) and focuses on scientific conferences.
As mentioned, the novelty of our work is that no other research analyzed specifically the correlation between the impact of a conference and the venue where it took place, as far as we know.
Previous works investigated what  are  the  main  factors  that  influence  conference participation decision-making (e.g. \cite{terzi2013international, Borghans2010868, Lee2019281}) and agree that, among the factors considered by the authors when deciding whether to attend a conference,
many concern the destination with its cultural and infrastructural potential. 
However, only a few recent works studied the relation between these factors and the impact of conferences (e.g. \cite{Goel2015115, lee2020author, Lee2019281}), and none of them analyzed the relation between the touristicity of a conference venue and quantitative indicators of the impact of research papers presented at conferences.

We also point out that this study provides a comprehensive view on the topic, being the size of our dataset much larger (i.e. at least two orders of magnitude greater, both in terms of conferences series and papers considered) than the ones used in similar analyses, as discussed in Section 2.



This paper is organized as follows: at first, we present some related work in Section~\ref{RelatedWork} to show the novelty of our research; we then moved to our methodology, detailing how we analyzed the data in Section~\ref{MethodsAndMaterials}; we show the results of our analysis in Section \ref{Results}, and we eventually conclude our work in Section~\ref{Conclusion}.


\section{Related Work}
\label{RelatedWork}

In the Computer Science field, 
conferences play a very important role. 
They are channels of paramount importance used by scientists to 
share their researches, as discussed in \cite{glanzel2006proceedings}, \cite{lisee2008conference} and \cite{vrettas2015conferences}.   

The importance of conferences in Computer Science is testified by quantitative indicators. For instance, Scopus 2020 Content Coverage Guide \cite{scopus2020} reports that 62.3\% of articles in Computing and Information Science were conference papers, while only about half (i.e. 32.8\%) were journal papers. In other disciplines like Chemical Sciences, Biological Sciences and Medical \& Health Sciences only 1.9\%, 2.7\% and 2.9\% were from conferences, respectively. 
In \cite{laender2008assessing} the authors compare the publications from 30 Computer Science graduate programs in North-American, Europe and Brazil and observe that the ratios of conference papers to journal articles were 2.9, 2.5 and 2.1, respectively. 
The empirical evidence supporting the impact of conferences in Computer Science is also discussed in \cite{Bar-Ilan2010809}. 

A lack of metrics supported by the citation databases is one of the main reasons that make hard to persuade non-computer scientists that conference papers have a value. 
Measures such as the acceptance rate of conferences are sometimes used as a proxy of quality, but cannot be compared with the Impact Factor of journals. 
Although Impact Factor is commonly calculated for journals only, recent studies developed proposals for a widely adoptable quality evaluation and ranking system for conferences, which could be equivalent to the Impact Factor (\cite{Li2018879,Loizides2017541}). 
Besides, \citeauthor{he2017usage} observe that conference papers, whose main objective is communication, have a stronger correlation with altmetrics than journal papers. 
Thus, altmetrics can be suitable to assess the impact of conference papers in a fast-moving research environment such that of Computer Science, as discussed in \cite{he2017usage}.
A recent study by \cite{yang2020proceedings} also reports that altmetrics indexes of conference papers in science fields are closely related to citations, reflecting the citation impact of proceedings papers.



One of the key aspect 
to explain the importance of conferences in the Computer Science domain 
is that, given the evolution speed of this discipline, conference papers are perceived as one of the most timely means for disseminating new ideas (\cite{Vardi20105,Eckmann2012617}).
In fact, unlike the cycle of publication of articles in journals, the time 
between the submission of an article in a conference and the editorial decision and publication is very short. 

Moreover, by offering various formats of presentations (e.g., discussions, presentations, posters, industry and position papers), conference open up opportunities to present cutting-edge ideas, even if less mature and sophisticated, bringing scientists together.
As a consequence, the attendees can disseminate and get instant feedbacks about their researches and ideas, and establish new collaborations triggered by on-site interactions, as described in \cite{Freyne2010124}.

Many studies investigated what are the main factors that influence conference participation decision-making. 
In \cite{terzi2013international} the authors analyze the answers of 123 academics and students to questionnaires based on the five-point Likert Scale, and discover that the destination, with its cultural and infrastructural potential, is one of the most important factors that determine the decision-making by researchers regarding their participation in conferences and exhibitions. 
The paper focuses on five evaluation criteria of conference location that have been derived from a literature review, and conclude that the most important ones are \textit{means of transport} (87.81\% of positive answers), \textit{safety/hygiene} (86.99\%) and \textit{use of international language} (83.74\%), followed by \textit{security} (77.24\%) and \textit{infrastructure} (73.98\%).


Similar findings are described in \cite{Borghans2010868}. Based on a sample of European labour economists, preferences are measured using the vignette approach where participants are asked to choose between hypothetical European Association of Labour Economists (EALE) conferences.
The results show that \textit{Keynote speakers} and \textit{conference location} are the most important attributes to decide for a conference, while the remaining attributes \textit{type of social event}, \textit{time of the year}, and \textit{conference venue} are less important for the decision.
The paper also shows that varying characteristics of a conference can influence both the overall attractiveness of a conference, but also influences the type of researchers interested in participation.


In \cite{factorsAffecting2008} the authors developed a reliable and valid five-factor scale for convention participation decision-making based on 558 survey responses.
The final measurement scale consists of five interrelated but unique dimensions of convention participation decision-making: \textit{destination stimuli}, \textit{professional and social networking opportunities}, \textit{educational opportunities}, \textit{safety and health situation}, and \textit{travelability}.


All these studies agree that, besides scientific motivations supported by bibliometric and quantitative indicators, most of the factors considered by the authors when deciding whether to attend a conference, thus determining its success, concern  the  destination,  with  its  cultural  and infrastructural potential.
However, only a few studies analyzed the relation between the touristicity of a conference venue and quantitative indicators of the impact of research papers presented at conferences.

A recent work by \cite{Lee2019281} examines 43,463 papers from 81 conference series in the Information Science and Computer Science fields and analyzes the contributions of conference related factors to the citation rates of the conference papers. 
The results of the study reports that two factors related to the attractiveness of the conference venue are able to predict the quality of presented papers, measured in terms of their citation rates: the \textit{seasonal accessibility of conferences} (e.g. when a conference is held during the middle of a school semester, faculty and student researchers may feel burdensome to cancel or skip classes, and therefore, if there is another alternative option that is held during the vacation season, the researchers may select that alternative) and the \textit{size of the conferences} in terms of the number of presented papers, which is a direct consequence of the venue attractiveness, especially for non top-rated conferences.
The regression results also illustrate that other aspects, such as longevity and names of the conference series, their acceptance rates, the content similarity of the presented papers at a conference, the degree of the authors' international collaborations and the records of the best paper awards at conferences are significantly predictive to the future citations of the conference papers. 

In \cite{Goel2015115} the authors analyze 63 networking conferences and about 39,000 conference papers published in these conferences during the years 2008 through 2012. 
An empirical study investigate the impact on the number of citations received by papers of three factors: conference acceptance rate, paper title and year of publication. The results show a correlation between the analyzed factors and citation count. However the authors acknowledge that there are several other factors (not considered in their empirical study) that may influence the citation count of conference papers by as much as 46\%.

Another study by \cite{lee2020author} analyze 12,237 papers published in 28 conference series in the Computer Science field.
The paper analyzes the predictive power of 21 factors related to the first author and to all authors on citation counts of conference papers. 
In particular, the work considers authors' academic performance, degree of collaboration and topological properties of their citation networks. 
The results show that the author-related factors have a significantly higher power for explaining conference paper impact in terms of citation counts than the first author-related factors. 
Among all author-related indicators, the average degree centrality of all authors is the most relevant factor for predicting future citations, followed by the average of all authors’ betweenness centrality and average citation rates of prior publications.

The works discussed in this section confirm the importance of conferences for specific research fields such as Computer Science.
However, as shown in Table \ref{tab:related}, previous studies considered only a limited subset of conferences in the Computer Science field, both in terms of conference series/number of papers and time span.
Moreover, most of them acknowledge that several other factors besides those traditionally considered by the bibliometric community should be considered to fully understand the impact of conferences, such as the conference destination with  its  cultural  and infrastructural potential.
However, at the best of our knowledge, no study has yet investigated in depth the relation between the touristicity of a conference location and quantitative indicators of the impact of research papers presented at conferences. 

\begin{table}[]
\centering
\renewcommand{\arraystretch}{1.5}
\begin{tabular}{l r r r}
\hline
\textbf{Work} & \textbf{N. conf. series} & \textbf{N. papers} & \textbf{Time Span}\\ \hline
\citeauthor{Goel2015115} (SEDE, \citeyear{Goel2015115}) & 63 & 38,895 & 5 years (2008-2012)\\ 
\citeauthor{Lee2019281} (Scientometrics, \citeyear{Lee2019281}) & 81 & 43,463 & 4 years (2009-2012)\\
\citeauthor{lee2020author} (Electronic Library, \citeyear{lee2020author})& 28 & 12,337 & 4 years (2009-2012)\\ 
This work & 3,838 & 2,581,564 & 33 years (1985-2017)\\ \hline
\end{tabular}
\renewcommand{\arraystretch}{1}
\caption{Comparison of the related work with our study.}
\label{tab:related}
\end{table}

\section{Methods and Materials}
\label{MethodsAndMaterials}

For the analyses described in this paper, we worked on publication citation data (Section~\ref{sec:citData}), quantifying citations for scientific conference papers, and integrating it with tourism information data (Section~\ref{sec:tourData}), characterizing the cities and countries that hosted the conferences each year. The data gathered and the software developed for our work are available in \cite{ThisArticleData} and in the GitHub repository\footnote{https://github.com/lbedogni/conference-touristicity}, respectively.

\begin{table}[t]
\centering
\renewcommand{\arraystretch}{1.5}
\begin{tabular}{p{1.2cm} p{2.1cm} l p{3cm} l l}
\hline
\textbf{authors} & \textbf{title}                      & \textbf{crossref} & \textbf{url}                           & \textbf{year} & \textbf{cit} \\
\hline
Mark   Harman    & Searching for better configurations & conf/sigsoft/2013 & db/conf/cdc/cdc2016\ldots & 2013          & 25           \\
\ldots                & \ldots                                   & \ldots                 & \ldots                                      & \ldots             & \ldots\\
\hline
\end{tabular}
\renewcommand{\arraystretch}{1}
\caption{Example of raw data fields}
\label{tab:exData}
\end{table}

\subsection{Citation data - conference impact.}
\label{sec:citData}

To quantify the impact of a conference, we started from the raw data extracted from the OpenCitations\footnote{https://opencitations.net/} website. 
First publishing open bibliographic data in 2010, the Initiative for Open Citations\footnote{https://i4oc.org} (I4OC) was born with the idea of promoting the release of open citation data, as discussed in \cite{shotton2013publishing}. 
As a result, now we have several millions of citation data openly available on the Web, important stakeholders supporting the movement, and several projects and studies (e.g. \cite{isse2019, Zhu20201097, scient2021}) leveraging the open citation data available online. 

In this work we used OpenCitations as the main datasource of bibliometric information for our analyses. 
The data is provided with records which identify each citing article and each cited article. Since we are focusing on the number of citation of a conference, we summed all the citations received by any article. We then merged the data with that available from DBLP\footnote{https://dblp.org/}, a free and publicly available Computer Science
bibliography repository started in 1993 at the University of Trier, which describes Computer Science conferences. We matched the two datasets on the conference at which an article was presented, and then summed all the citations of articles presented in any given conference. The resulting dataset contains complete citation data for scientific publications ranging from 1960 to the present day. Several variables 
are available, including: authors, title, booktitle, crossref, url (DBLP page link), editor, pages, publisher, series, year, cit (number of received citations).  
First of all, we projected the data on the variables 
that are relevant to our analysis: authors, title, crossref (containing a unique key for each edition of a conference), url, year, cit (see Table~\ref{tab:exData} for a sample publication entry). Then, we proceeded with data filtering/cleaning and we explicitly excluded from our working set:

\begin{itemize}
    \item publications not appearing in conference proceedings or without conference information (e.g., journals, technical reports, etc.);
    \item publications without citation information;
    \item recent publications from the last 2-3 years (staring from 2018), since they are too recent for citation analysis;
    \item publications from too early years (we grouped the data per year and noticed a crescent number of publications trend, we required at least 2500 publications for each analyzed year and thus we selected publications from 1985 onwards).
\end{itemize}

\begin{table}[t]
\centering
\renewcommand{\arraystretch}{1.5}
\begin{tabular}{p{2cm}|p{2cm} p{2cm} p{2cm} p{2cm}}
\hline
\textbf{conf}      & \textbf{totCit} & \textbf{avgCit} & \textbf{nPubs} & \textbf{nEditions} \\
\hline
\textbf{eurocrypt} & 52855           & 40.347          & 1310           & 36                 \\
\textbf{ismb}      & 10875           & 35.656          & 305            & 5                  \\
\textbf{…}         & …               & …               & …              & …\\
\hline
\end{tabular}
\renewcommand{\arraystretch}{1}
\caption{Citation data: high-level conference aggregate statistics}
\label{tab:confOverview}
\end{table}

\begin{table}[t]
\centering
\renewcommand{\arraystretch}{1.5}
\begin{tabular}{p{2cm}|p{0.8cm} p{1.7cm} p{1.7cm} p{1.7cm} p{1.7cm}}
\hline
\textbf{conf}     & \textbf{…} & \textbf{2014} & \textbf{2015} & \textbf{2016} & \textbf{2017} \\
\hline
\textbf{ACISicis} & …          & 53            & 79            & 148           & 89            \\
\textbf{ACMace}   & …          & 205           & 94            & 55            & 35            \\
\textbf{…}        & …          & …             & …             & …             & …\\
\hline
\end{tabular}
\renewcommand{\arraystretch}{1}
\caption{Citation data: temporal evolution of conference total citations per year}
\label{tab:totCitEvol}
\end{table}

We consequently obtained a working dataset of 2,581,564 entries. Starting from this data, we processed it in order to obtain different ``views'' and be able to later analyze it from different points of view (also by crossing such views with the tourism data which will be described in the next section):

\begin{itemize}
    \item \textbf{Conferences high-level aggregate statistics}: grouping by conference key (e.g., ``eurocrypt''), we computed total  (``totCit'') and average (``avgCit'') citation number, total number of publications (``nPubs'') and editions (``nEditions''), see Table \ref{tab:confOverview} for sample data. In particular, the total and average citation number are the two main indicators we will use to quantify the conference impact;
    \item \textbf{Conferences editions high-level  aggregate statistics}: same as last point, but with data grouped on conference edition (conference name and year, e.g., ``eurocrypt/2010'');
    \item \textbf{Temporal evolution data}: for each conference, the evolution through its different editions/years of the total number of citations (see Table \ref{tab:totCitEvol}), average number of citations, total number of publications;
    \item \textbf{Relative temporal evolution data (percentages)}: same as last point, but with data for each year expressed in percentage w.r.t. all editions/years (e.g., average citations for conference ``ismb'' for year XY divided by the average citations computed on all editions data, and so on for all years). This view enables a better glance at the relative variations of each monitored citation metric for each conference over time.
\end{itemize}

\subsection{Tourism data - conference venue ``touristicity''.}\label{sec:tourData}

\begin{table}[t]
\centering
\renewcommand{\arraystretch}{1.5}
\begin{tabular}{p{3cm}|p{2.5cm} p{2.5cm} p{2.5cm}}
\hline
\textbf{conf/year}      & \textbf{city} & \textbf{country} & \textbf{state} \\
\hline
\textbf{cscw/2000}      & Philadelphia  & USA              & PA             \\
\textbf{mobicom/2000}   & Boston        & USA              & MA             \\
\textbf{eurocrypt/2005} & Aarhus        & Denmark          & -              \\
\textbf{…}              & …             & …                & …\\
\hline
\end{tabular}
\renewcommand{\arraystretch}{1}
\caption{Tourism data: conference location information}
\label{tab:place}
\end{table}

\begin{table}[t]
\centering
\renewcommand{\arraystretch}{1.5}
\begin{tabular}{p{3cm}|p{1.33cm} p{1.33cm} p{1.33cm} p{1.33cm} p{1.33cm}}
\hline
\textbf{city}          & \textbf{totCit} & \textbf{avgCit} & \textbf{nEditions} & \textbf{SWP} & \textbf{TA} \\
\hline
\textbf{Beijing}       & 61913           & 168.70          & 368                & 179383       & 4           \\
\textbf{Paris}         & 62430           & 190.34          & 328                & 218583       & 17.56       \\
\textbf{San Francisco} & 82004           & 274.26          & 299                & 177857       & 2.9         \\
\textbf{…}             & …               & …               & …                  & …            & …\\
\hline
\end{tabular}
\renewcommand{\arraystretch}{1}
\caption{Citation/tourism data: aggregate citation and tourism information for each city}
\label{tab:city}
\end{table}

\begin{table}[t]
\centering
\renewcommand{\arraystretch}{1.5}
\begin{tabular}{p{1.6cm}|p{0.8cm} p{0.8cm} p{1.2cm} p{0.8cm} p{0.8cm} p{0.8cm} p{0.8cm} p{0.8cm}}
\hline
\textbf{country} & \textbf{totCit} & \textbf{avgCit} & \textbf{nEditions} & \textbf{TTCI} & \textbf{TSI} & \textbf{CRBT} & \textbf{GPI} & \textbf{SI} \\
\hline
\textbf{USA}     & 1539133         & 249.78          & 6167               & 5.3           & 6.6          & 4.7           & 4.9          & 21.5        \\
\textbf{China}   & 223458          & 130.45          & 1714               & 4.9           & 3.5          & 7             & 3.9          & 19.3        \\
\textbf{Germany} & 168601          & 124.61          & 1356               & 5.4           & 5.9          & 6.5           & 5.7          & 23.5        \\
\textbf{…}       & …               & …               & \textbf{…}         & …             & …            & …             & …            & …\\
\hline
\end{tabular}
\renewcommand{\arraystretch}{1}
\caption{Citation/tourism data: aggregate citation and tourism information for each country}
\label{tab:country}
\end{table}

To quantify the ``touristicity'' of a venue, we first of all extracted location information for the conferences of our dataset. In particular, we extracted the exact city, country and state for each edition of each conference (see Table \ref{tab:place} for an example). Then, we gathered and linked several statistics from external resources characterizing the extracted locations from different tourism-related points of view. More specifically, each city was associated to the following touristicity indicators:

\begin{itemize}
    \item \textbf{SWP} (Size of Wikipedia Page): this indicator represents our proposal to easily quantify the importance/attractiveness of the city and expresses the dimension (in bytes) of the associated Wikipedia page. The rationale is that the higher is the importance of a city the more information we expect to be available in its description;
    \item \textbf{TA} (Tourist Arrivals): this is the number of tourist arrivals (in millions) to each city as monitored in 2018 by Euromonitor\footnote{https://go.euromonitor.com/white-paper-travel-2018-100-cities}, for which we obtained all the relevant data and merged it with the conference data.
\end{itemize}

Moreover, for each country we extracted the following indicators from the latest Travel and Tourism competitiveness report\footnote{http://reports.weforum.org/travel-and-tourism-competitiveness-report-2019/}:


\begin{itemize}
\item \textbf{TTCI} (Travel \& Tourism Competitiveness Index), the most general index,  quantifying the whole set of factors and policies
that enable the sustainable development of the Travel
\& Tourism sector in a given country (range 1-7). In particular, the index is composed of 14 pillars organized into four categories: enabling environment
(business environment, safety and security, health and hygiene, human resources and labour market, ICT Readiness), T\&T policy and enabling conditions (prioritization of travel \& tourism, international openness, price competitiveness, environmental sustainability), infrastructures (air transport, ground and port, tourist service), natural and cultural resources (natural  resources, cultural resources and business travel);
\item \textbf{TSI} (Tourist Service Infrastructure), quantifying the availability of sufficient quality accommodation, resorts and entertainment facilities in a given country (range 1-7). The index is measured through the number of hotel rooms complemented by the extent of access to services such as car rentals and ATMs;
\item \textbf{CRBT} (Cultural Resources and Business Travel), measuring a country's cultural resources (range 1-7). Among the considered factors are the number of UNESCO cultural World Heritage sites, the number of large stadiums that can host significant sport or entertainment events, the number of online searches related to a country’s cultural resources and the number of international association meetings;
\item \textbf{GPI} (Ground and Port Infrastructure), measuring a country's  availability of efficient and accessible transportation to key business centres and tourist attractions (range 1-7). More specifically, the index considers the presence of extensive road and railroad network, proxied by road and railroad densities, as wells as roads, railroads, and ports infrastructure that meet international standards of comfort, security and modal efficiency.
\end{itemize}

Finally, by joining the above city and country tourism data with the citation data (total and average citation and number of editions, aggregated per city), we were able to obtain new combined views of our data (see Tables \ref{tab:city} and \ref{tab:country} for city and country, respectively).

\section{Results}
\label{Results}

In this section we discuss the results we have obtained, with respect to the variation in the citation number among the same conference in different years and in different places.

\begin{figure}
\centering
\begin{subfigure}[b]{0.65\textwidth}
\centering
\includegraphics[width=\textwidth]{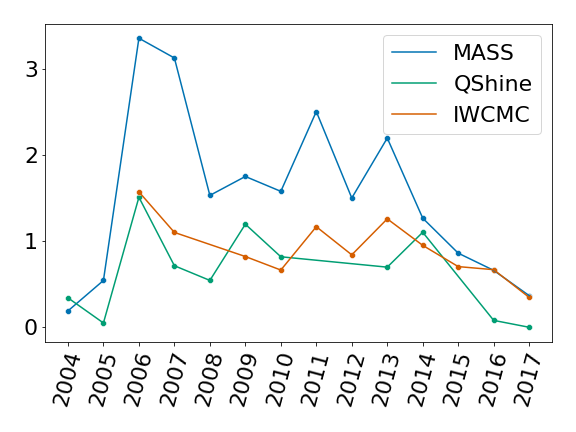}
\caption{Class 3}\label{fig:1a1}
\end{subfigure}
\begin{subfigure}[b]{0.65\textwidth}
\centering
\includegraphics[width=\textwidth]{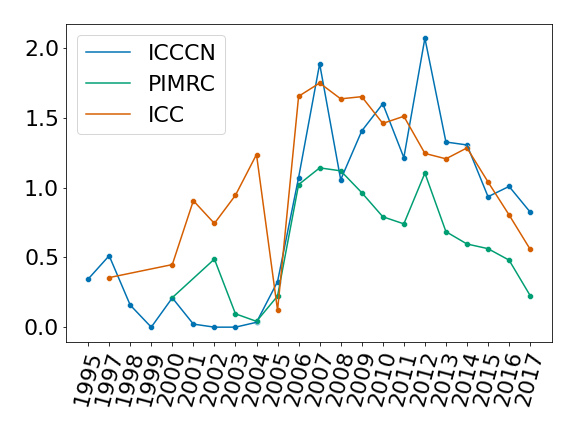}
\caption{Class 2}\label{fig:1a2}
\end{subfigure}
\begin{subfigure}[b]{0.65\textwidth}
\centering
\includegraphics[width=\textwidth]{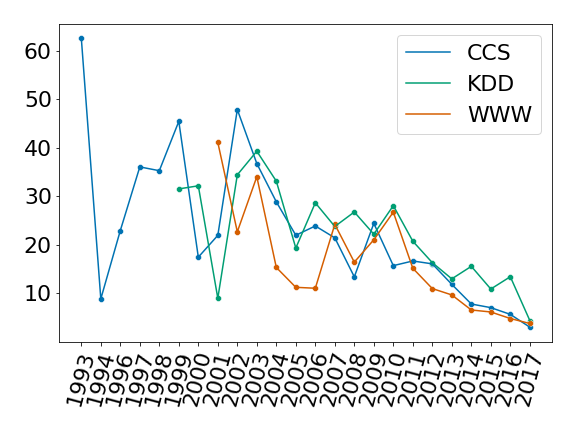}
\caption{Class 1}\label{fig:1a3}
\end{subfigure}
\caption{Examples of different conference with respect to the Average Citation number of each year.}\label{fig:example-avg}
\end{figure}

\begin{figure}
\centering
\begin{subfigure}[b]{0.65\textwidth}
\centering
\includegraphics[width=\textwidth]{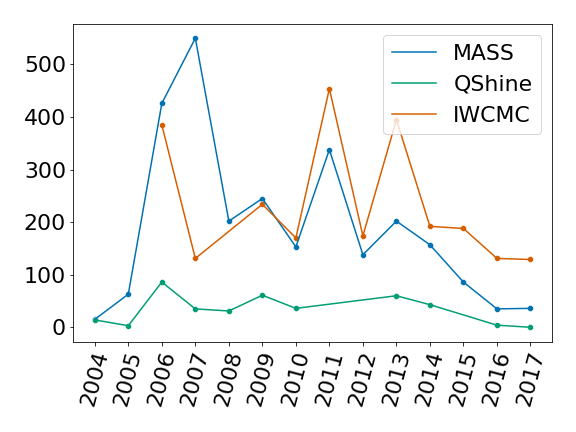}
\caption{Class 3}\label{fig:1a4}
\end{subfigure}
\begin{subfigure}[b]{0.65\textwidth}
\centering
\includegraphics[width=\textwidth]{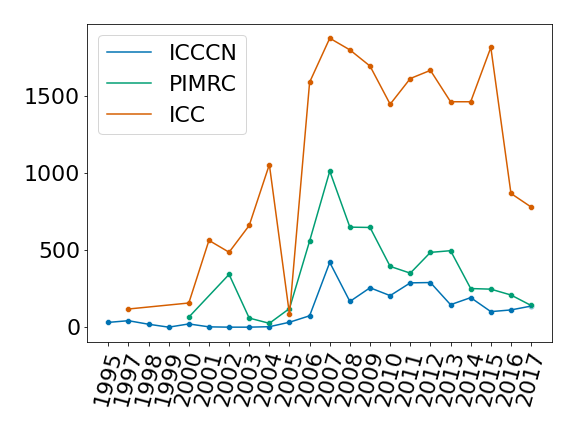}
\caption{Class 2}\label{fig:1a5}
\end{subfigure}
\begin{subfigure}[b]{0.65\textwidth}
\centering
\includegraphics[width=\textwidth]{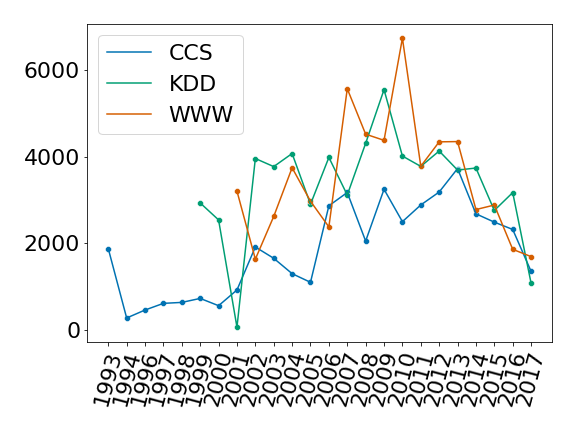}
\caption{Class 1}\label{fig:1a6}
\end{subfigure}
\caption{Examples of different conference with respect to the Total Citation number of each year.}\label{fig:example-tot}
\end{figure}

At first, we show some examples which highlight the differences in the number of citations, for the same conference series over different years. 
Figure~\ref{fig:example-avg} and Figure \ref{fig:example-tot} show the average citations for different conferences, for any given year. Each plot refers to different classes of conferences, according to the Italian Conference Ranking System\footnote{http://valutazione.unibas.it/gii-grin-scie-rating/}, which comprises several other international conference evaluations, such as the CORE conference ranking, Microsoft Academic, and LiveSHINE. We can see that, regardless of the class of the conference, both the average citation number and the total citation number vary a lot during different years, although for Class 1 conferences it is less evident. There is a global drop starting from around 2015, which is given to the fact that articles are new and have yet to collect a number of citations comparable to older articles. Although a slight variation between any year is normal, there are evident cases in which from one year to another the number of citations, and the average of it, vary even by $400\%$.

\begin{figure}
    \centering
    \includegraphics[width=0.65\textwidth]{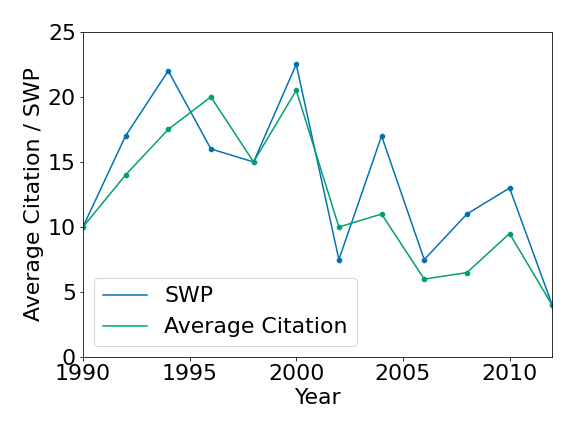}
    \caption{PPSN conference over the years}
    \label{fig:example-swp-avg}
\end{figure}
We now considered whether indicators such as those presented in Section \ref{sec:tourData} are correlated with the average citation number, meaning that better indicators should reflect a higher number of citations. In particular, we started by visually analyzing the trends focusing on one conference at a time. As an example, we show in Figure \ref{fig:example-swp-avg} the average citation number of the PPSN conference over the years, and the SWP indicator of the venue in which the conference took place on any given year. As it can be seen, the two lines follow a rather similar trend, which may indicate that indeed a correlation exists.

\begin{figure}
    \centering
    \includegraphics[width=1\textwidth]{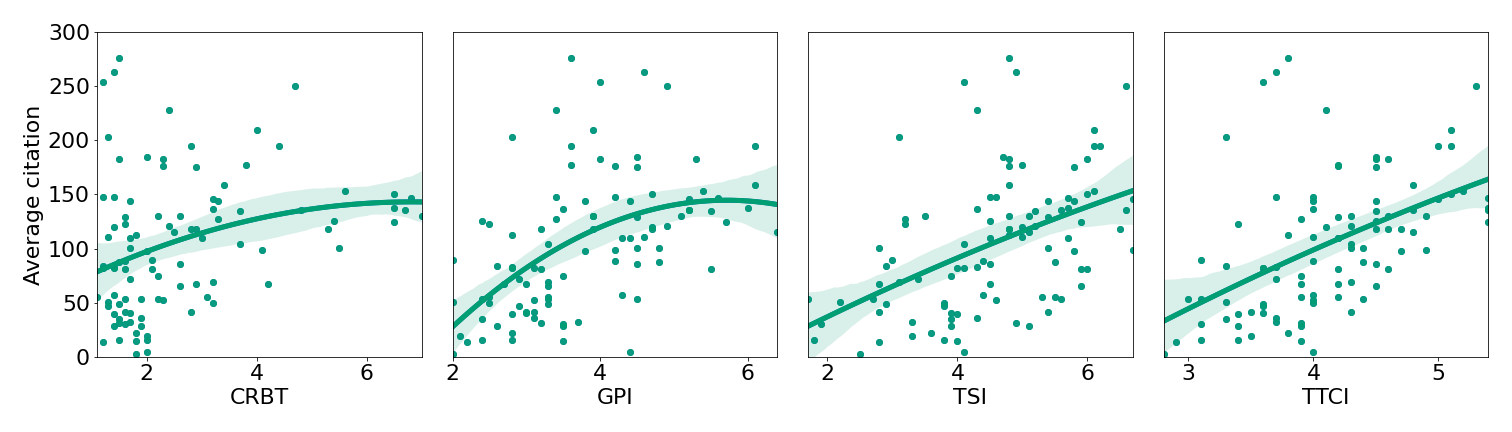}
    \caption{Average citations versus tourism indices}
    \label{fig:cit-vs-idx}
\end{figure}
Then, in order to generalize our findings, we considered all the conferences of our dataset and all the 4 country wide indicators presented in Section \ref{sec:tourData} (results are shown in Figure \ref{fig:cit-vs-idx}). Here, each point refers to a specific venue, and the average citation is the average of the citation number of any conference in that specific venue. On the x axis we show the different metric values for the touristicity metrics considered, and as it can be easily seen, the trend is linear for all four indicators, as also shown by the fitting curves. 

\begin{figure}
\centering
\begin{subfigure}[b]{0.65\textwidth}
\centering
\includegraphics[width=\textwidth]{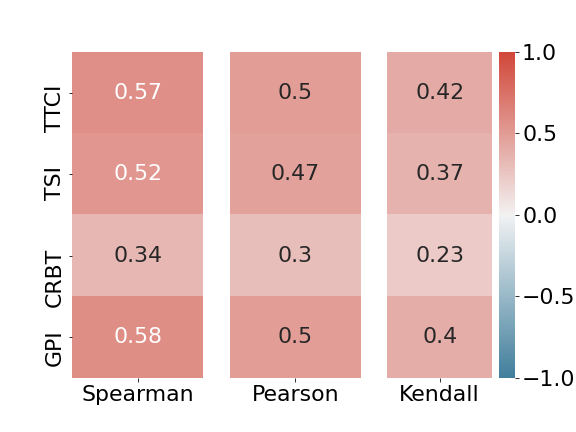}
\caption{Average Citations}\label{fig:corr_avg}
\end{subfigure}
\begin{subfigure}[b]{0.65\textwidth}
\centering
\includegraphics[width=\textwidth]{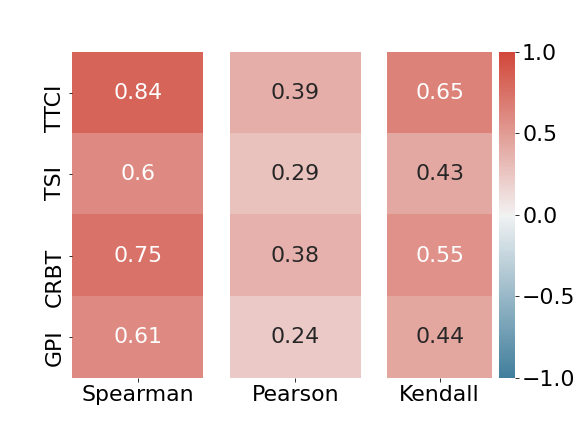}
\caption{Total Citations}\label{fig:corr_tot}
\end{subfigure}
\caption{Correlation between the different indices and the Average Citation number (\figurename~\ref{fig:corr_avg}) and the Total Citation number (\figurename~\ref{fig:corr_tot}) for Country wide indices.}
\label{fig:corr}
\end{figure}

\begin{figure}
\centering
\begin{subfigure}[b]{0.65\textwidth}
\centering
\includegraphics[width=\textwidth]{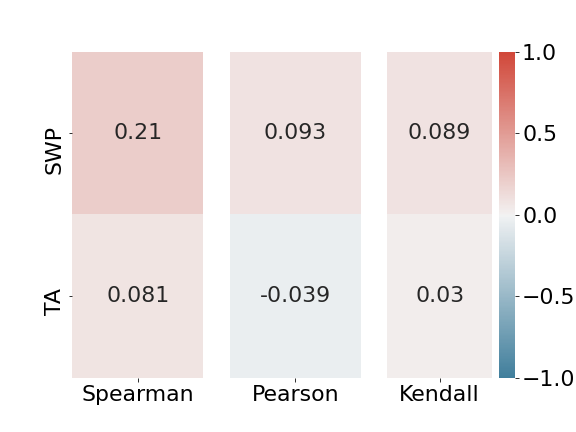}
\caption{Average Citations}\label{fig:corr_avg_city}
\end{subfigure}
\begin{subfigure}[b]{0.65\textwidth}
\centering
\includegraphics[width=\textwidth]{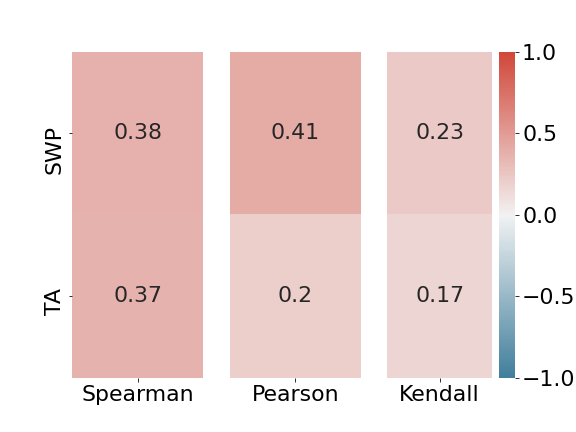}
\caption{Total Citations}\label{fig:corr_tot_city}
\end{subfigure}
\caption{Correlation between the different indices and the Average Citation number (\figurename~\ref{fig:corr_avg_city}) and the Total Citation number (\figurename~\ref{fig:corr_tot_city}) for City wide indices.}
\label{fig:corr_city}
\end{figure}






Figure \ref{fig:corr} shows instead the results of the correlation between the Average Citation number (Figure \ref{fig:corr_avg}) and the touristicity metrics, and the Total Citation number (Figure \ref{fig:corr_tot}) and the same metrics. In these two figures, all the touristicity indicators are related to the country in which the conference took place. We started our analysis from the data related to all the conferences divided by year, and correlated all of them together with the touristicity indicators. For sake of readability, we do not report the p-values for each correlation values, which however are all below $1^{-5}$, therefore representing a trusted correlation measure. We have performed the tests using three different correlation measures, namely Spearman, Pearson and Kendall. What it can be seen is that according to Spearman there is a rather strong correlation with both the Average and the Total Citation number and various indices which represent the touristic attractiveness of a city in which a conference took place. Pearson is the one which shows less correlation, though for some indices it highlights a correlation of about $0.5$. Kendall shows instead good correlation for a higher number of metrics, both for the Average and the Total Citation number. In any case, it is evident that it exists a correlation between the citation number of a conference and the city in which it took place. 


Figure \ref{fig:corr_city} shows instead indices related to the attractiveness of a city instead of a country. Here we can see that although in some cases there are quite interesting correlations, values are much lower compared to Figure \ref{fig:corr}. Therefore, we can say that according to our analysis, it exists a stronger correlation when considering the country in which the conference is organized, rather than the specific city. We also note that this particular aspect is one of the main future work on this topic.

\section{Discussion and concluding remarks}
\label{Conclusion}

In this paper we have presented a study related to the correlations between the attractiveness of the scientific conferences and the ``touristicity'' of their venues.


According to our original research questions, we have shown that a correlation exists between the number of the citations of the conference publications and some touristic indicators of the conference venue.
So we can conclude the paper answering the two research questions.
\begin{itemize}
    \item \textbf{RA1}: Yes, there is a correlation between the impact of a scientific conference and the venue where it takes place, and the correlation is stronger when considering country-wide indicators;
    \item \textbf{RA2}: The two touristic indexes that are more involved in the correlation are the TTCI, Travel \& Tourism Competitiveness Index, and the TSI, which is the Tourist Service Infrastructure. The high correlation of the TTCI, being it the most comprehensive index among the ones considered, corroborates our general findings, while the particular presence of the TSI attests the specific importance of the presence of tourist/accommodation facilities in a given country.
\end{itemize}

The above indicators present an almost linear correlation with the number of citations, and this also shows that country-wide indexes are more correlated with the impact than city-wide indexes; this can be explained because the chance of visiting a touristic region goes beyond the visit of a single city.

We are aware of the two main limitations of our work: (i) the use of citations to evaluate the attractiveness of the conferences and (ii) the difficulty to formally define the touristic attractiveness of a venue. 

In our future work we aim at overcoming the former limitation by analyzing the number of the submissions instead of the number of citations; this should better represent the interest in submitting a paper to any conference, depending on the year hence on the organization location. This will also be useful to validate the assumption that a large number of submissions implies more citations. Moreover, we will consider: a) the definition of other touristicity indicators by extracting data from further sources, trying to quantify the impact of specific factors to the overall venue attractiveness; b) the possibility of further expanding the dataset.
We also remark that the results open up new possibilities, such as allowing both conference organizers and authors to estimate in advance the impact of conferences, thus supporting them in their decisions.

\section{Declarations}
\label{Declarations}

\noindent \textit{Funding:} no funding was received for conducting this study.

\noindent \textit{Conflicts of interest/Competing interests:} the authors have no relevant financial or non-financial interests to disclose.

\noindent \textit{Availability of data and material:} https://doi.org/10.5281/zenodo.4734182.

\noindent \textit{Code availability:} https://github.com/lbedogni/conference-touristicity

\noindent \textit{Authors' contributions: } all authors contributed equally to this research.





\bibliographystyle{spbasic_updated}      
\bibliography{biblio}   

\end{document}